\documentclass[reprint,superscriptaddress,amsmath,amssymb,prb]{revtex4-1}

\usepackage{lipsum}
\usepackage{graphicx}
\usepackage{dcolumn}
\usepackage{bm}
\usepackage{hyperref}
\usepackage{tabularx}
\usepackage[version=4]{mhchem}
\usepackage{physics}
\usepackage{ulem}

\begin{document}

\title{Comment on ``Low-frequency lattice phonons in halide perovskites explain high defect tolerance toward electron-hole recombination"}

\author{Sunghyun Kim}
\affiliation{Department of Materials, Imperial College London, Exhibition Road, London SW7 2AZ, UK}

\author{Aron Walsh}
\affiliation{Department of Materials, Imperial College London, Exhibition Road, London SW7 2AZ, UK}
\affiliation{Department of Materials Science and Engineering, Yonsei University, Seoul 03722, Korea}

\date{\today}

\begin{abstract}
Halide perovskites exhibit slow rates of non-radiative electron-hole recombination upon illumination. Chu \textit{et al.} [\textit{Sci. Adv.} \textbf{6} 7, eaaw7453 (2020)] use the results of first-principles simulations to argue that this arises from the nature of the crystal vibrations and leads to a breakdown of Shockley-Read-Hall theory. We highlight flaws in their methodology and analysis of carrier capture by point defects in crystalline semiconductors.
\end{abstract}

\maketitle
The theory of carrier capture and recombination in semiconductors has a rich history. 
Early developments include the statistical model of Shockley and Read\cite{shockley1952statistics} that could explain the  measurements of Hall\cite{hall1952electron}, as well as the spectroscopic models of Huang and Rhys.\cite{huangrhys} 
These were complemented by more rigorous theories over the subsequent decades that have been extensively reviewed.\cite{stoneham1981non}
Often ``multi-phonon emission" is highlighted, which is linked to the structural change that occurs upon charge trapping.\cite{henry1977nonradiative,das2020deep}
There has been a recent renaissance in the theory led by the first-principles formalisms including those of Alkauskas \textit{et al.}\cite{alkauskas2014first} and 
Shi \textit{et al.},\cite{shi2015comparative} which have been applied to a range of semiconductor hosts.

The defect tolerance of hybrid perovskites to non-radiative electron-hole recombination in solar cells is well established in the literature.
\cite{yin2014unusual,brandt2015identifying,walsh2017instilling}
In their article, Chu \textit{et al.}\cite{chu2020low} explore the behaviour of five point defects in methylammonium lead iodide (MAPI) using first-principles non-adiabatic molecular dynamics (NAMD). 
We argue that their methodology is insufficient to explain the non-radiative electron-hole recombination rates of halide perovskite semiconductors and does not support a breakdown of conventional models. 

\subsection*{Resilience of Shockley-Read-Hall theory}
Shockley-Read-Hall (SRH) theory\cite{shockley1952statistics,hall1952electron} describes the statistics of sequential carrier capture by trap levels in semiconducting materials based on the principle of detailed balance.
The balance of electron and hole capture is determined by the steady state population of the charge states of defects,
which is a function of carrier concentration, capture cross-section, and thermodynamic charge-transition (trap) level.
A deep trap may have a large or small capture cross section for an electron or hole, but the same statistics still apply.

Under conditions of high illumination intensity, two-carrier (radiative) and three-carrier (Auger) recombination processes often dominate, which does not contradict SRH theory.\cite{nelson2003physics}
These other processes have recently been explored for MAPI.\cite{zhang2019first}
While Chu \textit{et al.} argue for the ``breakdown of SRH theory'', they do not actually consider the relevant capture processes and the corresponding recombination statistics to support their claim.

Significant weight is given to analysis of the electronic density of states (DOS) in Ref. \onlinecite{chu2020low}. 
It is well established that the single-particle DOS is a poor descriptor of the quasi-particle optical or thermal trap levels associated with a defect in an extended solid. 
An ionisation process, e.g. 
\begin{equation}
\ce{
D^0 <=>[\Delta E]
D^+ + e^-}
\end{equation}
includes electronic and structural relaxation of the final charge state [$D^+$], as well as the energy of an electron far from the defect [$e^-$]. 
This is why a $\Delta$-SCF type-approach is used as a standard protocol for defect calculations,
making them comparable to a range of deep level spectroscopies.
Defects that feature a shallow DOS may indeed exhibit deep trap states.

In Ref. \onlinecite{chu2020low}, all defects are modelled in  \textit{neutral} charge states.
These are not the ground-state configurations\cite{yin2014unusual} and indeed the electrostatics of charged defects is a primary driving force for carrier capture.
For example, the iodine vacancy introduces a resonance level into the conduction band of MAPI. 
In an undoped material, it will exist as \ce{V_I^+} with the electron donated to the reservoir of conduction electrons. 
An appropriate initial charge state is essential to describe trap-mediated recombination events.

\begin{table*}[ht!]
\begin{tabular}[t]{ccc}
          \hline
    \textbf{Pseudo wave function} & & \textbf{All-electron wave function} \\
    \begin{tabular}{c|cccccc}
             \hline
          & VBM-3 & VBM & VBM & VBM & CBM & CBM+1  \\
          \hline
        VBM-3 & 1.25  & 0    & 0    & 0    & 0.08 & 0    \\
        VBM   & 0     & 1.07 & 0    & 0    & 0    & 0   \\
        VBM   & 0     & 0    & 1.07 & 0    & 0    & 0.01 \\
        VBM   & 0     & 0    & 0    & 1.07 & 0    & 0.01 \\
        CBM   & 0.08  & 0    & 0    & 0    & 1.45 & 0    \\
        CBM+1 & 0     & 0    & 0.01 & 0.01 & 0    & 1.05  \\
            \hline
    \end{tabular} &      &
    \begin{tabular}{c|cccccc}
    \hline
          & VBM-3 & VBM & VBM & VBM & CBM & CBM+1  \\
          \hline
        VBM-3 & 1 & 0 & 0 & 0 & 0 & 0 \\
        VBM   & 0 & 1 & 0 & 0 & 0 & 0 \\
        VBM   & 0 & 0 & 1 & 0 & 0 & 0 \\
        VBM   & 0 & 0 & 0 & 1 & 0 & 0 \\
        CBM   & 0 & 0 & 0 & 0 & 1 & 0 \\
        CBM+1 & 0 & 0 & 0 & 0 & 0 & 1 \\
            \hline
    \end{tabular}
\end{tabular}
\caption{Calculated overlap between the Kohn-Sham wavefunctions of \ce{GaAs} at the $\Gamma$ point. 
Calculations were performed with the PAW method implemented in \textsc{VASP} as reported in Ref. \onlinecite{kim2019anharmonic}. 
The core part of the all-electron wave function is evaluated using \textsc{pawpyseed}.\cite{bystrom2019pawpyseed}
The valence band maximum (VBM) is triply degenerate, while the conduction band minimum (CBM) is nondegenerate.}
\label{tab:pawvsae}
\end{table*}

The most basic interaction between a charge carrier and a defect in a dielectric host is electrostatic. 
This effect is often accounted for by the Sommerfeld Factor ($s$).\cite{stoneham1981non}
The Coulomb interaction between a charge carrier $q$ and point defect $Q$ can be attractive ($Z = Q/q < 0$) or repulsive ($Z = Q/q > 0$). For the attractive case, the relationship is:
\begin{equation}
s = 4|Z| \sqrt{\frac{\pi E_R}{k_BT}}
\end{equation}
where $E_R$ is the effective Rydberg energy 
\begin{equation}
E_R = \frac{m^*q^4}{2 \hbar^2 \epsilon_0^2}
\end{equation}
There are two consequences for halide perovskites. 
Firstly, the halide anion ($Q = -1$) ensures that $Q$ will remain small compared to chalcogenides ($Q = -2$)  or pnictides  ($Q = -3$).
Secondly, large static dielectric constants ($\epsilon_0 > 20$)
ensures that $E_R$ remains small. 
Together, a singly charged defect in MAPI (e.g. \ce{V_I}$^+$) is screened by 10$\times$ compared to a triply charged defect (e.g. \ce{V_{As}}$^{+++}$) in GaAs.
Thus, long-range electrostatic and dielectric effects  will play an important role in the carrier capture and defect tolerance of MAPI,
but are not considered in Ref. \onlinecite{chu2020low}.

\subsection*{Unphysical non-radiative recombination}
Theories of non-radiative carrier capture 
stem from Landau-Zener theory
that describes the probability of transition following a non-adiabatic level crossing.
In Ref. \onlinecite{chu2020low} and recent publications\cite{Zhang:2019gb, Chu:2020is} fast ``direct'' non-radiative electron-hole recombination has been predicted in pristine crystalline semiconductors.
The 1.5 eV band gap should act as a substantial barrier that prohibits band-to-band non-radiative recombination. The simulations imply a crossing of valence and conduction bands; however, such band gap collapse is unphysical at room temperature.
For MAPI, Chu \textit{et al.} indeed showed that the electronic eigenvalues of the valence band maximum (VBM) and conduction band minimum (CBM) do not vary significantly during the molecular dynamics simulations.

We argue that the non-adiabatic coupling (NAC) matrix elements $d_{jk}$ between the valence and conduction bands must be significantly overestimated in Ref. \onlinecite{chu2020low}.
The approach results in exaggerated non-radiative recombination in pristine systems. 
For example, carrier lifetimes have been observed to exceed 100 $\mu$s 
in MAPI single crystals,
yet they calculate ``20\% direct non-radiative recombination'' after just 2 ns.
Due to a poor description of the perfect crystal, defects can then appear to hinder recombination in the simulation time scale.
Indeed, one central conclusion is that an iodine interstitial can \textit{reduce} non-radiative recombination. 
Similar claims have been made about the lead vacancy.\cite{he2018lead}

NAMD  is powerful for its description of molecular excitations and chemical transformations.\cite{tapavicza2013ab}
In the \textsc{Hefei-NAMD}\cite{hefei_namd} implementation, the matrix elements are calculated using the time-derivative of the overlap between two Kohn-Sham wave functions:
\begin{equation}
    d_{jk} = \mel{\psi_j}{\frac{\partial}{\partial t}}{\psi_k}
\end{equation}
The underlying code \textsc{VASP} uses the projector-augmented-wave (PAW) method,\cite{Blochl1994} and the pseudo wave functions at the $\Gamma$-point (from WAVECAR) are used to calculate $d_{jk}$. 
However, these wave functions are neither eigenstates of the Hamiltonian nor orthogonal.\cite{Blochl1994}
We illustrate for the case of zinc-blende GaAs in Table \ref{tab:pawvsae}.
Description of the overlap matrix requires transformation to all-electron wave functions are  orthonormal.
The values obtained from pseudo wave functions are not simply rescaled, but feature off-diagonal elements. The spurious nature of the coupling must impact the time evolution of NAMD. In addition, the band-edge states of semiconductors will be poorly described due to  $\Gamma$-point sampling of the Brillouin zone.

Even if the proper matrix elements were employed, the kinetics of trap-mediated recombination are first-order and limited by the capture of the minority carrier. 
For a positive trap, the excess electronic energy is emitted as heat ($\hbar\omega$) at each sequential \textit{discrete} capture events:
 \begin{equation}
 \ce{
 D^+ <=>[h\nu]
 D^+ + e^- + h^+ <=>[-\hbar\omega] D^0 + h^+ <=>[-\hbar\omega] D^+}
 \end{equation}
 In contrast, NAMD simulates a bimolecular electron--hole excitation of a defective unit cell ($\sim 10^{21}$ cm$^{-3}$ excitons) and monitors the \textit{continuous} fractional decay ($\delta$) of an excited state after a finite simulation time (2 ns), i.e.
 \begin{equation}
 \ce{
 D^0 + [e^- + h^+]  -> D^0 + (1-\delta)[e^- + h^+] 
 }
 \end{equation}
To describe realistic SRH non-radiative recombination in a crystalline semiconductor using a real time approach, a large supercell and long timescales would be required, likely beyond the means of present simulation capability.

\bibliography{defects}

\begin{thebibliography}{22}%
\makeatletter
\providecommand \@ifxundefined [1]{%
 \@ifx{#1\undefined}
}%
\providecommand \@ifnum [1]{%
 \ifnum #1\expandafter \@firstoftwo
 \else \expandafter \@secondoftwo
 \fi
}%
\providecommand \@ifx [1]{%
 \ifx #1\expandafter \@firstoftwo
 \else \expandafter \@secondoftwo
 \fi
}%
\providecommand \natexlab [1]{#1}%
\providecommand \enquote  [1]{``#1''}%
\providecommand \bibnamefont  [1]{#1}%
\providecommand \bibfnamefont [1]{#1}%
\providecommand \citenamefont [1]{#1}%
\providecommand \href@noop [0]{\@secondoftwo}%
\providecommand \href [0]{\begingroup \@sanitize@url \@href}%
\providecommand \@href[1]{\@@startlink{#1}\@@href}%
\providecommand \@@href[1]{\endgroup#1\@@endlink}%
\providecommand \@sanitize@url [0]{\catcode `\\12\catcode `\$12\catcode
  `\&12\catcode `\#12\catcode `\^12\catcode `\_12\catcode `\%12\relax}%
\providecommand \@@startlink[1]{}%
\providecommand \@@endlink[0]{}%
\providecommand \url  [0]{\begingroup\@sanitize@url \@url }%
\providecommand \@url [1]{\endgroup\@href {#1}{\urlprefix }}%
\providecommand \urlprefix  [0]{URL }%
\providecommand \Eprint [0]{\href }%
\providecommand \doibase [0]{http://dx.doi.org/}%
\providecommand \selectlanguage [0]{\@gobble}%
\providecommand \bibinfo  [0]{\@secondoftwo}%
\providecommand \bibfield  [0]{\@secondoftwo}%
\providecommand \translation [1]{[#1]}%
\providecommand \BibitemOpen [0]{}%
\providecommand \bibitemStop [0]{}%
\providecommand \bibitemNoStop [0]{.\EOS\space}%
\providecommand \EOS [0]{\spacefactor3000\relax}%
\providecommand \BibitemShut  [1]{\csname bibitem#1\endcsname}%
\let\auto@bib@innerbib\@empty
\bibitem [{\citenamefont {Shockley}\ and\ \citenamefont
  {Read}(1952)}]{shockley1952statistics}%
  \BibitemOpen
  \bibfield  {author} {\bibinfo {author} {\bibfnamefont {W.}~\bibnamefont
  {Shockley}}\ and\ \bibinfo {author} {\bibfnamefont {W.}~\bibnamefont
  {Read}},\ }\href@noop {} {\bibfield  {journal} {\bibinfo  {journal} {Phys.
  Rev.}\ }\textbf {\bibinfo {volume} {87}},\ \bibinfo {pages} {835} (\bibinfo
  {year} {1952})}\BibitemShut {NoStop}%
\bibitem [{\citenamefont {Hall}(1952)}]{hall1952electron}%
  \BibitemOpen
  \bibfield  {author} {\bibinfo {author} {\bibfnamefont {R.~N.}\ \bibnamefont
  {Hall}},\ }\href@noop {} {\bibfield  {journal} {\bibinfo  {journal} {Phys.
  Rev.}\ }\textbf {\bibinfo {volume} {87}},\ \bibinfo {pages} {387} (\bibinfo
  {year} {1952})}\BibitemShut {NoStop}%
\bibitem [{\citenamefont {Huang}\ and\ \citenamefont {Rhys}(1950)}]{huangrhys}%
  \BibitemOpen
  \bibfield  {author} {\bibinfo {author} {\bibfnamefont {K.}~\bibnamefont
  {Huang}}\ and\ \bibinfo {author} {\bibfnamefont {A.}~\bibnamefont {Rhys}},\
  }\href@noop {} {\bibfield  {journal} {\bibinfo  {journal} {Proc. Roy. Soc.
  A}\ }\textbf {\bibinfo {volume} {204}},\ \bibinfo {pages} {406} (\bibinfo
  {year} {1950})}\BibitemShut {NoStop}%
\bibitem [{\citenamefont {Stoneham}(1981)}]{stoneham1981non}%
  \BibitemOpen
  \bibfield  {author} {\bibinfo {author} {\bibfnamefont {A.}~\bibnamefont
  {Stoneham}},\ }\href@noop {} {\bibfield  {journal} {\bibinfo  {journal} {Rep.
  Prog. Phys.}\ }\textbf {\bibinfo {volume} {44}},\ \bibinfo {pages} {1251}
  (\bibinfo {year} {1981})}\BibitemShut {NoStop}%
\bibitem [{\citenamefont {Henry}\ and\ \citenamefont
  {Lang}(1977)}]{henry1977nonradiative}%
  \BibitemOpen
  \bibfield  {author} {\bibinfo {author} {\bibfnamefont {C.}~\bibnamefont
  {Henry}}\ and\ \bibinfo {author} {\bibfnamefont {D.~V.}\ \bibnamefont
  {Lang}},\ }\href@noop {} {\bibfield  {journal} {\bibinfo  {journal} {Phys.
  Rev. B}\ }\textbf {\bibinfo {volume} {15}},\ \bibinfo {pages} {989} (\bibinfo
  {year} {1977})}\BibitemShut {NoStop}%
\bibitem [{\citenamefont {Das}\ \emph {et~al.}(2020)\citenamefont {Das},
  \citenamefont {Aguilera}, \citenamefont {Rau},\ and\ \citenamefont
  {Kirchartz}}]{das2020deep}%
  \BibitemOpen
  \bibfield  {author} {\bibinfo {author} {\bibfnamefont {B.}~\bibnamefont
  {Das}}, \bibinfo {author} {\bibfnamefont {I.}~\bibnamefont {Aguilera}},
  \bibinfo {author} {\bibfnamefont {U.}~\bibnamefont {Rau}}, \ and\ \bibinfo
  {author} {\bibfnamefont {T.}~\bibnamefont {Kirchartz}},\ }\href@noop {}
  {\bibfield  {journal} {\bibinfo  {journal} {Phys. Rev. Mater.}\ }\textbf
  {\bibinfo {volume} {4}},\ \bibinfo {pages} {024602} (\bibinfo {year}
  {2020})}\BibitemShut {NoStop}%
\bibitem [{\citenamefont {Alkauskas}\ \emph {et~al.}(2014)\citenamefont
  {Alkauskas}, \citenamefont {Yan},\ and\ \citenamefont {Van~de
  Walle}}]{alkauskas2014first}%
  \BibitemOpen
  \bibfield  {author} {\bibinfo {author} {\bibfnamefont {A.}~\bibnamefont
  {Alkauskas}}, \bibinfo {author} {\bibfnamefont {Q.}~\bibnamefont {Yan}}, \
  and\ \bibinfo {author} {\bibfnamefont {C.~G.}\ \bibnamefont {Van~de Walle}},\
  }\href@noop {} {\bibfield  {journal} {\bibinfo  {journal} {Phys. Rev. B}\
  }\textbf {\bibinfo {volume} {90}},\ \bibinfo {pages} {075202} (\bibinfo
  {year} {2014})}\BibitemShut {NoStop}%
\bibitem [{\citenamefont {Shi}\ \emph {et~al.}(2015)\citenamefont {Shi},
  \citenamefont {Xu},\ and\ \citenamefont {Wang}}]{shi2015comparative}%
  \BibitemOpen
  \bibfield  {author} {\bibinfo {author} {\bibfnamefont {L.}~\bibnamefont
  {Shi}}, \bibinfo {author} {\bibfnamefont {K.}~\bibnamefont {Xu}}, \ and\
  \bibinfo {author} {\bibfnamefont {L.-W.}\ \bibnamefont {Wang}},\ }\href@noop
  {} {\bibfield  {journal} {\bibinfo  {journal} {Phys. Rev. B}\ }\textbf
  {\bibinfo {volume} {91}},\ \bibinfo {pages} {205315} (\bibinfo {year}
  {2015})}\BibitemShut {NoStop}%
\bibitem [{\citenamefont {Yin}\ \emph {et~al.}(2014)\citenamefont {Yin},
  \citenamefont {Shi},\ and\ \citenamefont {Yan}}]{yin2014unusual}%
  \BibitemOpen
  \bibfield  {author} {\bibinfo {author} {\bibfnamefont {W.-J.}\ \bibnamefont
  {Yin}}, \bibinfo {author} {\bibfnamefont {T.}~\bibnamefont {Shi}}, \ and\
  \bibinfo {author} {\bibfnamefont {Y.}~\bibnamefont {Yan}},\ }\href@noop {}
  {\bibfield  {journal} {\bibinfo  {journal} {App. Phys. Lett.}\ }\textbf
  {\bibinfo {volume} {104}},\ \bibinfo {pages} {063903} (\bibinfo {year}
  {2014})}\BibitemShut {NoStop}%
\bibitem [{\citenamefont {Brandt}\ \emph {et~al.}(2015)\citenamefont {Brandt},
  \citenamefont {Stevanovi{\'c}}, \citenamefont {Ginley},\ and\ \citenamefont
  {Buonassisi}}]{brandt2015identifying}%
  \BibitemOpen
  \bibfield  {author} {\bibinfo {author} {\bibfnamefont {R.~E.}\ \bibnamefont
  {Brandt}}, \bibinfo {author} {\bibfnamefont {V.}~\bibnamefont
  {Stevanovi{\'c}}}, \bibinfo {author} {\bibfnamefont {D.~S.}\ \bibnamefont
  {Ginley}}, \ and\ \bibinfo {author} {\bibfnamefont {T.}~\bibnamefont
  {Buonassisi}},\ }\href@noop {} {\bibfield  {journal} {\bibinfo  {journal}
  {MRS Commun.}\ }\textbf {\bibinfo {volume} {5}},\ \bibinfo {pages} {265}
  (\bibinfo {year} {2015})}\BibitemShut {NoStop}%
\bibitem [{\citenamefont {Walsh}\ and\ \citenamefont
  {Zunger}(2017)}]{walsh2017instilling}%
  \BibitemOpen
  \bibfield  {author} {\bibinfo {author} {\bibfnamefont {A.}~\bibnamefont
  {Walsh}}\ and\ \bibinfo {author} {\bibfnamefont {A.}~\bibnamefont {Zunger}},\
  }\href@noop {} {\bibfield  {journal} {\bibinfo  {journal} {Nature Mater.}\
  }\textbf {\bibinfo {volume} {16}},\ \bibinfo {pages} {964} (\bibinfo {year}
  {2017})}\BibitemShut {NoStop}%
\bibitem [{\citenamefont {Chu}\ \emph {et~al.}(2020{\natexlab{a}})\citenamefont
  {Chu}, \citenamefont {Zheng}, \citenamefont {Prezhdo}, \citenamefont {Zhao},\
  and\ \citenamefont {Saidi}}]{chu2020low}%
  \BibitemOpen
  \bibfield  {author} {\bibinfo {author} {\bibfnamefont {W.}~\bibnamefont
  {Chu}}, \bibinfo {author} {\bibfnamefont {Q.}~\bibnamefont {Zheng}}, \bibinfo
  {author} {\bibfnamefont {O.~V.}\ \bibnamefont {Prezhdo}}, \bibinfo {author}
  {\bibfnamefont {J.}~\bibnamefont {Zhao}}, \ and\ \bibinfo {author}
  {\bibfnamefont {W.~A.}\ \bibnamefont {Saidi}},\ }\href@noop {} {\bibfield
  {journal} {\bibinfo  {journal} {Science Adv.}\ }\textbf {\bibinfo {volume}
  {6}},\ \bibinfo {pages} {eaaw7453} (\bibinfo {year}
  {2020}{\natexlab{a}})}\BibitemShut {NoStop}%
\bibitem [{\citenamefont {Nelson}(2003)}]{nelson2003physics}%
  \BibitemOpen
  \bibfield  {author} {\bibinfo {author} {\bibfnamefont {J.}~\bibnamefont
  {Nelson}},\ }\href@noop {} {\emph {\bibinfo {title} {Physics of Solar
  Cells}}}\ (\bibinfo  {publisher} {World Scientific Publishing Company},\
  \bibinfo {year} {2003})\BibitemShut {NoStop}%
\bibitem [{\citenamefont {Zhang}\ \emph
  {et~al.}(2019{\natexlab{a}})\citenamefont {Zhang}, \citenamefont {Shen},\
  and\ \citenamefont {Van~de Walle}}]{zhang2019first}%
  \BibitemOpen
  \bibfield  {author} {\bibinfo {author} {\bibfnamefont {X.}~\bibnamefont
  {Zhang}}, \bibinfo {author} {\bibfnamefont {J.-X.}\ \bibnamefont {Shen}}, \
  and\ \bibinfo {author} {\bibfnamefont {C.~G.}\ \bibnamefont {Van~de Walle}},\
  }\href@noop {} {\bibfield  {journal} {\bibinfo  {journal} {Adv. Energy
  Mater.}\ }\textbf {\bibinfo {volume} {100}},\ \bibinfo {pages} {1902830}
  (\bibinfo {year} {2019}{\natexlab{a}})}\BibitemShut {NoStop}%
\bibitem [{\citenamefont {Kim}\ \emph {et~al.}(2019)\citenamefont {Kim},
  \citenamefont {Hood},\ and\ \citenamefont {Walsh}}]{kim2019anharmonic}%
  \BibitemOpen
  \bibfield  {author} {\bibinfo {author} {\bibfnamefont {S.}~\bibnamefont
  {Kim}}, \bibinfo {author} {\bibfnamefont {S.~N.}\ \bibnamefont {Hood}}, \
  and\ \bibinfo {author} {\bibfnamefont {A.}~\bibnamefont {Walsh}},\
  }\href@noop {} {\bibfield  {journal} {\bibinfo  {journal} {Phys. Rev. B}\
  }\textbf {\bibinfo {volume} {100}},\ \bibinfo {pages} {041202} (\bibinfo
  {year} {2019})}\BibitemShut {NoStop}%
\bibitem [{\citenamefont {Bystrom}\ \emph {et~al.}(2019)\citenamefont
  {Bystrom}, \citenamefont {Broberg}, \citenamefont {Dwaraknath}, \citenamefont
  {Persson},\ and\ \citenamefont {Asta}}]{bystrom2019pawpyseed}%
  \BibitemOpen
  \bibfield  {author} {\bibinfo {author} {\bibfnamefont {K.}~\bibnamefont
  {Bystrom}}, \bibinfo {author} {\bibfnamefont {D.}~\bibnamefont {Broberg}},
  \bibinfo {author} {\bibfnamefont {S.}~\bibnamefont {Dwaraknath}}, \bibinfo
  {author} {\bibfnamefont {K.~A.}\ \bibnamefont {Persson}}, \ and\ \bibinfo
  {author} {\bibfnamefont {M.}~\bibnamefont {Asta}},\ }\href@noop {} {\bibfield
   {journal} {\bibinfo  {journal} {arXiv preprint arXiv:1904.11572}\ }
  (\bibinfo {year} {2019})}\BibitemShut {NoStop}%
\bibitem [{\citenamefont {Zhang}\ \emph
  {et~al.}(2019{\natexlab{b}})\citenamefont {Zhang}, \citenamefont {Chu},
  \citenamefont {Zheng}, \citenamefont {Benderskii}, \citenamefont {Prezhdo},\
  and\ \citenamefont {Zhao}}]{Zhang:2019gb}%
  \BibitemOpen
  \bibfield  {author} {\bibinfo {author} {\bibfnamefont {L.}~\bibnamefont
  {Zhang}}, \bibinfo {author} {\bibfnamefont {W.}~\bibnamefont {Chu}}, \bibinfo
  {author} {\bibfnamefont {Q.}~\bibnamefont {Zheng}}, \bibinfo {author}
  {\bibfnamefont {A.~V.}\ \bibnamefont {Benderskii}}, \bibinfo {author}
  {\bibfnamefont {O.~V.}\ \bibnamefont {Prezhdo}}, \ and\ \bibinfo {author}
  {\bibfnamefont {J.}~\bibnamefont {Zhao}},\ }\href@noop {} {\bibfield
  {journal} {\bibinfo  {journal} {J Phys. Chem. Lett.}\ }\textbf {\bibinfo
  {volume} {10}},\ \bibinfo {pages} {6151} (\bibinfo {year}
  {2019}{\natexlab{b}})}\BibitemShut {NoStop}%
\bibitem [{\citenamefont {Chu}\ \emph {et~al.}(2020{\natexlab{b}})\citenamefont
  {Chu}, \citenamefont {Saidi}, \citenamefont {Zhao},\ and\ \citenamefont
  {Prezhdo}}]{Chu:2020is}%
  \BibitemOpen
  \bibfield  {author} {\bibinfo {author} {\bibfnamefont {W.}~\bibnamefont
  {Chu}}, \bibinfo {author} {\bibfnamefont {W.~A.}\ \bibnamefont {Saidi}},
  \bibinfo {author} {\bibfnamefont {J.}~\bibnamefont {Zhao}}, \ and\ \bibinfo
  {author} {\bibfnamefont {O.~V.}\ \bibnamefont {Prezhdo}},\ }\href {\doibase
  10.1002/anie.201915702} {\bibfield  {journal} {\bibinfo  {journal} {Angew.
  Chem. Int. Ed.}\ } (\bibinfo {year} {2020}{\natexlab{b}}),\
  10.1002/anie.201915702}\BibitemShut {NoStop}%
\bibitem [{\citenamefont {He}\ and\ \citenamefont {Long}(2018)}]{he2018lead}%
  \BibitemOpen
  \bibfield  {author} {\bibinfo {author} {\bibfnamefont {J.}~\bibnamefont
  {He}}\ and\ \bibinfo {author} {\bibfnamefont {R.}~\bibnamefont {Long}},\
  }\href@noop {} {\bibfield  {journal} {\bibinfo  {journal} {J. Phys. Chem.
  Lett.}\ }\textbf {\bibinfo {volume} {9}},\ \bibinfo {pages} {6489} (\bibinfo
  {year} {2018})}\BibitemShut {NoStop}%
\bibitem [{\citenamefont {Tapavicza}\ \emph {et~al.}(2013)\citenamefont
  {Tapavicza}, \citenamefont {Bellchambers}, \citenamefont {Vincent},\ and\
  \citenamefont {Furche}}]{tapavicza2013ab}%
  \BibitemOpen
  \bibfield  {author} {\bibinfo {author} {\bibfnamefont {E.}~\bibnamefont
  {Tapavicza}}, \bibinfo {author} {\bibfnamefont {G.~D.}\ \bibnamefont
  {Bellchambers}}, \bibinfo {author} {\bibfnamefont {J.~C.}\ \bibnamefont
  {Vincent}}, \ and\ \bibinfo {author} {\bibfnamefont {F.}~\bibnamefont
  {Furche}},\ }\href@noop {} {\bibfield  {journal} {\bibinfo  {journal} {Phys.
  Chem. Chem. Phys.}\ }\textbf {\bibinfo {volume} {15}},\ \bibinfo {pages}
  {18336} (\bibinfo {year} {2013})}\BibitemShut {NoStop}%
\bibitem [{hef()}]{hefei_namd}%
  \BibitemOpen
  \href@noop {} {}\bibinfo {howpublished}
  {\url{https://github.com/QijingZheng/Hefei-NAMD}}\BibitemShut {NoStop}%
\bibitem [{\citenamefont {Bl{\"o}chl}(1994)}]{Blochl1994}%
  \BibitemOpen
  \bibfield  {author} {\bibinfo {author} {\bibfnamefont {P.~E.}\ \bibnamefont
  {Bl{\"o}chl}},\ }\href@noop {} {\bibfield  {journal} {\bibinfo  {journal}
  {Phys. Rev. B}\ }\textbf {\bibinfo {volume} {50}},\ \bibinfo {pages} {17953}
  (\bibinfo {year} {1994})}\BibitemShut {NoStop}%
\end{thebibliography}%

\end{document}